\documentclass[slac_one]{revtex4}
\usepackage{graphicx}
\usepackage{epsfig}
\usepackage[dvips]{color}
\usepackage{amsmath,amssymb}
\usepackage{bm}
\usepackage{fancyhdr}
\newcommand{\EP}{\mbox{$e^+$}}
\newcommand{\EM}{\mbox{$e^-$}}
\newcommand{\EPEM}{\mbox{$e^+e^{-}$}}
\newcommand{\EMEM}{\mbox{$e^-e^-$}}
\newcommand{\GG}{\mbox{$\gamma\gamma$}}
\newcommand{\GE}{\mbox{$\gamma e$}}

\newcommand{\LGG}{\mbox{$L_{\gamma\gamma}$}}

\newcommand{\MKM}{\mbox{$\mu$m}}

\newcommand{\be}{\begin{equation}}
\newcommand{\ee}{\end{equation}}
\newcommand{\bc}{\begin{center}}
\newcommand{\ec}{\end{center}}
\newcommand{\bi}{\begin{itemize}}
\newcommand{\ei}{\end{itemize}}

\begin{document}

\title{{\small{2005 International Linear Collider Workshop - Stanford,
U.S.A.}}\\ 
\vspace{12pt} Comparison of photon colliders based on $ \bm{e^-e^-}$ and 
$\bm {e^+e^-}$ beams} 

%

\author{V.~I. Telnov}
\affiliation{Institute of Nuclear Physics, 630090 Novosibirsk, Russia}

\begin{abstract}
   At photon colliders (\GG,\GE) high energy photons are produced by
   Compton scattering of laser light off the high energy electrons (or
   positrons) at a linear collider. At first sight, photon colliders
   based on \EPEM\ or \EMEM\ primary beams have similar properties and
   therefore for convenience one can use \EPEM\ beams both for \EPEM\
   and \GG, \GE\ modes of operation. Below we compare these options
   and show that \EMEM\ beams are much better (mandatory) because in
   the \EPEM\ case low energy background $\GG\to$ hadrons is much
   higher and \EPEM\ annihilation reactions present a very serious
   background for \GG\ processes.
\end{abstract}

\maketitle
\thispagestyle{fancy}
\section{INTRODUCTION} 
   A photon collider is a very natural supplement to \EPEM\ linear
colliders.  Using Compton scattering of laser photons off high energy
electron (positron) beams one can obtain colliding \GG, \GE\ beams with
the energy and luminosity close to those in \EPEM\
collisions~\cite{GKST81,GKST83,GKST84,TESLATDR}.  It is usually
assumed that photon colliders are based on \EMEM\ beams because
production of electrons with a high degree of polarization is much easier. 
There are several other arguments in favor of \EMEM\ beams that were known to
photon collider experts but not emphasized because there was no such need.

  Recently, designers of the interaction region raised the
question about using \EPEM\ beams for the photon collider at the ILC because
some schemes of the final focus do not allow easy switching between
\EPEM\ and \EMEM\ modes and the disruption angles with \EPEM\ beams
should be smaller, which would make  beam removal somewhat easier.

Below I present strong arguments in favor of using \EMEM\ beams. In summary,
\bi
\item the study of \EMEM\ interactions is a part of the ILC physics
  program, and so \EMEM\ beams are necessary in any case;
\item a photon collider based on \EMEM\ beams is much better due to
    larger luminosity and much lower backgrounds from \EPEM\
    annihilation  and low energy $\GG\to$ hadrons reactions.
 \ei
\section{LUMINOSITY}
  If both \EP\ and \EM\ beams are prepared in the same damping rings,
  then their emittances  and the geometric (without beam
  collision effects) \EPEM\ and \EMEM\ luminosities are  equal.
  The \GG\ luminosity in the high energy part of the luminosity
  spectrum, which presents the main interest for physics, is simply
  proportional to the geometric luminosity. However, there is some
  difference in the degree of the polarization: for electrons
  $2\lambda_{e^-}=85$~\% \cite{clendenin} (determined by photo-guns),
  for positrons $2\lambda_{e^+}\sim 50$\% \cite{flottmann} (determined
  by the positron production scheme based on the process of \EPEM\
  pair production by polarized photons), where $\lambda_e$ is the
  helicity ($|\lambda_e<1/2|$).  The \GG\ luminosity at the high
  energy peak of the luminosity spectrum is just proportional to the
  product of photon spectra~\cite{GKST84} 
\be
  \left(\frac{dL}{dW_{\GG}}\right)_{peak} \propto
  \left[x+1+\frac{1}{x+1}-2P_c\lambda_i \frac{x(x+2)}{(x+1} \right]
  \left[x+1+\frac{1}{x+1}-2P_c\lambda_j \frac{x(x+2)}{(x+1} \right]\,,
\ee 
where $x=4E_0 \omega_0/m_e^2 c^4$, $E_0$ is the electron beam energy,
$\omega_0$ the laser photon energy and $P_c$ is the helicity of laser
photons.  

For the optimum of $x=4.8$ (the threshold for \EPEM\ pair
production at the conversion region) and $P_c=-1$ (gives maximum
luminosity in the high energy peak), we get

$$\left(\frac{dL}{dW_{\GG}}\right)_{\EPEM} \biggm/
\left(\frac{dL}{dW_{\GG}}\right)_{\EMEM} \sim 0.82 \;\;\;\; \mbox{for}\;\;\;\;
2\lambda_{e^+}=0.5$$
\vspace{-0.8cm}
\be
\hspace*{6.1cm} 0.58   \;\;\;\; \mbox{for unpolarized $e^+$}.
\ee

\begin{figure}[!htb]
\vspace{-0.6cm}
\centering
\includegraphics[width=160mm]{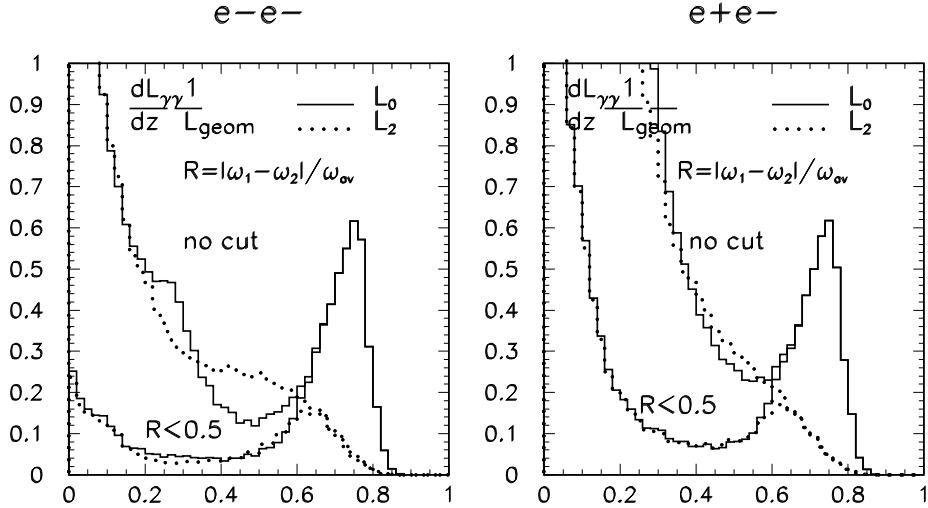}
\vspace{-1.6cm}
\caption{\GG\ luminosity spectra for \EPEM\ and \EMEM\ initial
colliding beams with and without a cut on the longitudinal momentum;
$2\lambda P_c = -0.85$ for both beams; indices 0, 2 are total helicities of
colliding photons.}
\label{fig1}
\end{figure}

The reduction of the high energy \GG\ luminosity for the \EPEM\ case
is quite noticeable. However, more informative is the comparison of \GG\ luminosity
spectra.  The \GG\ luminosity spectra for the TESLA TDR beams parameters
at $2E_0=500$ GeV in the \EPEM\ and \EMEM\ cases obtained by the 
simulation code~\cite{TEL95,TESLATDR} are presented in
Fig.~\ref{fig1}. The main results are summarized in Table~\ref{tab1}.

\begin{table}[!htb]
\vspace{-0.6cm}
\caption{Beam parameters and luminosities at the photon collider based
  on \EPEM\ and \EMEM\ beams, $2E_0=500$ GeV, the transverse beam
  emittances are the same as in the TESLA TDR~\cite{TESLATDR}} 
\bc
  {\renewcommand{\arraystretch}{1} \setlength{\tabcolsep}{3.8mm}
\begin{tabular}{l  c c} \hline
& \EMEM & \EPEM \\ \hline
$N/10^{10}$ & 2 & 2 \\
$\sigma_z$, mm & 0.3 & 0.3 \\
$\sigma_x$, nm & 88 & 88 \\
$\sigma_y$, nm & 4.3 & 4.3 \\
$L_{\mathrm{geom}},10^{35}$ & 1.2 & 1.2 \\
$\LGG(z>0.65)/L_{\mathrm{geom}}$ &{\color{blue} 0.1} & {\color{blue}0.1} \\ 
$\LGG(tot)/L_{\mathrm{geom}}$ & 0.92  & {\color{red} 5.6 } \\ 
$L_{\EMEM}/L_{\mathrm{geom}}$ & 0.006 &  --- \\
$L_{\EPEM}(z>0.65)/L_{\mathrm{geom}}$ & --- &   {\color{red} 0.062} \\
$L_{\EPEM}(tot)/L_{\mathrm{geom}}$ & --- & {\color{red} 0.24} \\ 
\end{tabular} \\[-0.5cm]
}
\ec
\label{tab1}
\end{table}
We see that the \GG\ luminosities in the high energy peak are equal
for \EMEM\ and \EPEM\ because we assumed similar $e^+$ and $e^-$
properties, including polarizations. It is about 10\% of the geometric
luminosity. However, there are two serious disadvantages of \EPEM\
beams: \bi
\item The total (mainly low energy) \GG\ luminosity with \EPEM\ beams
  is larger than with \EMEM\ beam by a factor of 6. The corresponding
  number of $\GG\to$ hadron reactions per one beam collision with \EPEM\
  beams will be about ten per bunch crossing!
\item In the case of  \EPEM\ beams,  $L(\EPEM)$) and $L(\GG)$ in the high
  energy region are comparable, and so  it will be difficult to separate
  \EPEM\ and \GG\ reactions due to  similar final states.
 \ei
 \section{LUMINOSITY MEASUREMENT}

The primary calibration processes for \GG\ collisions are $\GG\to
\EPEM\ \mbox{or}\;\; \mu^+\mu^-$.  At the photon collider with \EPEM\
beams similar final states will be produced in \EPEM\
annihilation. The residual \EPEM\ luminosity is comparable to the high
energy \GG\ luminosity, so the number of pairs produced in \EPEM\ and
\GG\ reactions will also be similar. It is possible, but very
difficult, to extract the \GG\ luminosity using differences in the
angular distributions of pairs in \EMEM\ and \GG\ processes.

  Note that at the photon collider with \EMEM\ beams the \EPEM\ luminosity
is also non-zero due to positron production at the conversion region
(Sect.~\ref{disr}) and at the interaction region (coherent pair
creation \cite{TESLATDR}), but it  will be much smaller than in the
case of \EPEM\ beams.

\section{PHYSICS BACKGROUNDS}

At a photon collider based on \EPEM\ beams, the residual \EPEM\
luminosity leads to a high rate of annihilation reactions $\EPEM\to X$,
which look very similar to $\GG\to X$ and present a serious problem for
analysis of two-photon processes.

Such a problem exists for study of charged pair production in \GG\
  collisions.  Only a combined analysis of \EPEM\ and \GG\ reactions by
  fitting their angular distributions allows to separate these processes at
  the price of a significant loss of the statistical accuracy.

For $\GG\to H \to b\bar{b}$ (the main decay mode for the light Higgs
boson), the most serious background at a \EMEM\ photon collider is the
QED process $\GG\to b\bar{b}$, which can be suppressed by a proper  choice of
photon polarizations. Indeed, $\sigma (\GG\to H) \propto
1+\lambda_{\gamma,1}\lambda_{\gamma,2}$, while $\sigma (\GG\to
b\bar{b}) \propto 1-\lambda_{1}\lambda_{2}$, where
photon helicities $\lambda_i$  are close to 100\% in the high energy part of
the luminosity spectrum.  

At a \EPEM\ based photon collider, additional background for the
Higgs arises from the reaction $\EPEM\to b\bar{b}$. It is not suppressed by
the beam polarizations because $\sigma_{e^+e^-}\propto
(1-\lambda_e\lambda_{e^+})\sim (1-0.8\cdot0.6)\sim 0.5$ and
$\sigma(\EPEM\to b\bar{b}) \sim \sigma(\GG\to b\bar{b})$. 

Note that at the photon collider the horizontal beam size should be as
small as possible, which causes large beamstrahlung losses for initial
charged particles during beam collisions. As a  result, for the
photon collider with \EPEM\ beams the residual \EPEM\ luminosity
spectrum is very broad and overlaps with the \GG\ luminosity spectrum.
    
\section{DISRUPTION ANGLES} {\label{disr}}

 At a photon collider with \EMEM\ initial beams, the maximum
 disruption angle is about 10--12 mrad~\cite{TESLATDR}. Large angles
 are caused by the repulsion of the low energy (after multiple Compton
 scattering) electrons off the opposing electron beam. One can expect
 that for \EPEM\ initial beams, the disruption angles are much smaller
 because particles attract to the opposing beam. However, even here
 \EPEM\ beams have no preferences due to the following reason.

In the $e\to\gamma$ conversion region, a high energy photon can
produce a \EPEM\ pair in collision with a laser photon. The threshold
for this process is $x=4.8$~\cite{GKST83,TESLATDR}. For $2E_0=500$ GeV and
a laser wavelength of $\lambda=1.06$ \MKM, $x \approx 4.5$, just
somewhat below the threshold. For higher energies (above the
threshold), this process has a cross section comparable to the
Compton one. As a result, after the conversion region the beam will
contain particles of both signs.  
Moreover, even at the parameter $x$ several times below the threshold
value of $x=4.8$, the high energy photon at the conversion region can  with
a rather high probability produce a \EPEM\ pair in a  collision with several
laser photons. 
So, the disrupted beams contain particles of both signs, so the
maximum disruption angles at \EPEM\ and \EMEM--based photon
colliders are approximately equal.

\section{CONCLUSION}

   Comparison of photon colliders based on \EPEM\ and \EMEM\ beams
shows that 

\bi

\item the \GG\ luminosity (in the high energy region) in the case of \EPEM\
beams will be  smaller by a factor of 1.2--1.8  depending on the polarization
degree of positrons;
\item the  low energy \GG\ luminosity with \EPEM\ beams
is greater by a factor of 6 for considered beam parameters.  The
corresponding hadronic background (about 10 events per bunch crossing)
leads to degradation of the energy and mass resolutions for most 
physics processes;

\item at a photon collider with \EPEM\  beams, the residual
\EPEM\ luminosity is comparable to the \GG\ luminosity and reactions
look very similar. This causes a very serious problem for distinguishing
\GG\ reactions and worsens the statistical accuracy.  Due to the same
problem, it is difficult to measure the \GG\ luminosity spectrum using
\EPEM\ or $\mu^+\mu^-$ pairs.

\item the maximum disruption angles at a photon collider with \EPEM\ and
\EMEM\ beams are similar due to intensive \EPEM\ pair production in the
conversion region.

\ei In summary, a photon collider with \EMEM\ initial beams has many
advantages compared to that based on \EPEM\ beams. I would put it even
stronger: \EPEM\ beams are absolutely unsuitable for photon colliders due to
very serious problems with identification of \GG\ and \EPEM\ reactions.

\end{document}